\documentclass[preprint]{elsarticle}

\usepackage{lineno,hyperref,graphicx,natbib}
\usepackage{float}
\biboptions{authoryear}

\usepackage{longtable}
\usepackage{booktabs}
\makeatletter
\@ifpackageloaded{subfig}{}{\usepackage{subfig}}
\@ifpackageloaded{caption}{}{\usepackage{caption}}
\AtBeginDocument{%

}
\AtBeginDocument{%

}
\@ifpackageloaded{float}{}{\usepackage{float}}
\floatstyle{ruled}
\@ifundefined{c@chapter}{\newfloat{codelisting}{h}{lop}}{\newfloat{codelisting}{h}{lop}[chapter]}
\floatname{codelisting}{Listing}

\makeatother

\setkeys{Gin}{width=\maxwidth,height=\maxheight,keepaspectratio}

\begin{document}

\begin{frontmatter}

\title{Dynamic foot morphology explained through 4D scanning and shape modeling}

\author[CUAero]{Abhishektha Boppana\corref{cor1}}
\ead{abhishektha@colorado.edu}

\author[CUAero]{Allison P. Anderson}

\address[CUAero]{Ann and H.J. Smead Department of Aerospace Engineering Sciences, University of Colorado Boulder, USA}

\cortext[cor1]{Corresponding Author}

\begin{abstract}
A detailed understanding of foot morphology can enable the design of more comfortable and better fitting footwear. However, foot morphology varies widely within the population, and changes dynamically during the loading of stance phase. This study presents a parametric statistical shape model from 4D foot scans to capture both the inter- and intra-individual variability in foot morphology. Thirty subjects walked on a treadmill while 4D scans of their right foot were taken at 90 frames-per-second during stance phase. Each subject's height, weight, foot length, foot width, arch length, and sex were also recorded. The 4D scans were all registered to a common high-quality foot scan, and a principal component analysis was done on all processed 4D scans. Elastic-net linear regression models were built to predict the principal component scores, which were then inverse transformed into 4D scans. The best performing model was selected with leave-one-out cross-validation. The chosen model was predicts foot morphology across stance phase with a root-mean squared error of 5.2 \(\pm\) 2.0 mm. This study shows that statistical shape modeling can be used to predict dynamic changes in foot morphology across the population. The model can be used to investigate and improve foot-footwear interaction, allowing for better fitting and more comfortable footwear.
\end{abstract}


\end{frontmatter}

\begin{center}\rule{0.5\linewidth}{0.5pt}\end{center}

\begin{center}\rule{0.5\linewidth}{0.5pt}\end{center}

\hypertarget{sec:intro}{%
\section{Introduction}\label{sec:intro}}

Foot shape is known to be highly variable throughout the population, including by sex \citep{Wunderlich2001, Krauss2008, Krauss2010}, age \citep{Tomassoni2014}, and weight \citep{Price2016}.
This variability is often not captured in footwear sizing, as current footwear fitting standards only use foot length, foot width, and arch length to fit to standardized shoe sizes \citep{ASTM2017}.
Furthermore, footwear is commonly designed around lasts, shoe molds that are sized and shaped by each manufacturer with no common standard, leading to variability in footwear shapes and sizes \citep{Jurca2013, Wannop2019}.
Such variability can make it hard for consumers to find a proper fit, resulting in users having to wear ill-fitting footwear with suboptimal comfort \citep{Dobson2018b}.
Footwear comfort has shown benefits in increasing running performance \citep{Luo2009} and reducing the risk of movement-related injury \citep{Mundermann2001a}, and is often the number one \citep{Martinez-Martinez2017} factor for consumers to select footwear.
Footwear should therefore be properly fit to a wide population range in order to be successful.

However, because the current methodology of designing footwear relies on using static lasts, this assumes that the foot consists of rigid segments.
This fails to account for dynamic changes in foot morphology, especially when the foot is being loaded during gait.
Assumptions of rigid foot segments during foot loading have shown inaccuracies in estimation of ankle joint mechanics \citep{Zelik2018, Kessler2020}, suggesting intra-foot motion as the foot is loaded \citep{Lundgren2008, Wolf2008}.
Evidence suggests that foot loading affects linear foot measurements, such as when transitioning from sitting to standing \citep{Xiong2009, Oladipo2008} or during the stance phase of gait \citep{Kouchi2009, Barisch-Fritz2014, Grau2018}.
The dynamically changing measurements suggest morphological changes occurring, all of which may not be captured in static linear and circumferential measurements.
Thus, it becomes difficult to characterize the wide variety of foot shapes across not only a large population, but within individuals as their foot goes through loading scenarios such as gait.

Statistical shape models (SSMs) can explain morphological differences across populations by identifying shape modes which account for variance from the mean foot,.
These have been developed for whole-body digital human modeling applications to study population and individual variance in body shape \citep{Allen2003, Anguelov2005, Reed2014, Park2015a, Park2017}.
Parametric SSMs are extensions which use correlations between subject anthropometric data and SSM deformations to help predict body shape for new individuals in the population \citep{Park2015a, Park2017}.

SSMs have recently been applied to characterize static foot shape across a population \citep{Conrad2019} and recognize foot-shape deviations \citep{Stankovic2020}.
The aforementioned efforts to capture foot measurement changes over the gait cycle did capture 4D foot images \citep{Barisch-Fritz2014, Grau2018}, but these efforts were not translated into a SSM.
All the previously developed systems were also based on a catwalk, requiring subjects to correctly hit the scanning area for a successful data capture, which may not be representative of natural cadence.

The development of the DynaMo software \citep{Boppana2019} for the Intel RealSense D415 Depth Cameras (Intel, Santa Clara CA) allowed a 4D scanning system to be set around a treadmill, where subjects can maintain a natural cadence.
This system captures the majority of the foot's dorsal surface, but does not allow for the capture of the foot's plantar surface.
4D scans are captured at 90 fps, enabling a detailed evaluation of foot morphology changes during loading and unloading.
This study outlines the development of a parametric SSM, derived from scans captured with this system.
The parametric SSM can characterize and predict dynamic foot morphology at specific points during stance phase across the subject population.
We hypothesize that there will be significant changes in foot morphology across the dorsal surface of the foot throughout the gait cycle.
We also hypothesize that these changes will be predictable from the subject demographics of our population.

\hypertarget{methods}{%
\section{Methods}\label{methods}}

\hypertarget{subjects}{%
\subsection{Subjects}\label{subjects}}

A total of 30 healthy subjects (15 men and 15 women, ages 23.1 \(\pm\) 3.7) participated in this study.
Subjects were recruited in a stratified sample into one of six groups (5 subjects per group) to maximize variance in population foot length.
Height was used as the grouping factor since height is well correlated to foot length \citep{Giles1991}. The general population may not know offhand their exact foot length, and shoe size varies by manufacturer and does not correspond directly to foot length \citep{Jurca2013, Wannop2019}. Groups consisted of 5th-35th, 35th-65th, and 65th-95th height percentiles for each sex.
Height percentile values were taken from the ANSUR II survey \citep{Gordon2014} and converted to imperial units as it was expected most subjects would report their height in imperial units.
Population recruitment groups are summarized in tbl.~\ref{tbl:groups}.

Prior to recruitment, subjects completed a prescreening survey to ensure they were adequately healthy by the American College of Sports Medicine guidelines\citep{Riebe2015}, and between the ages of 18-65.
Subjects provided their sex and height, and were only enrolled in the study if their population group was not fully enrolled.

\hypertarget{experimental-procedures}{%
\subsection{Experimental Procedures}\label{experimental-procedures}}

The experimental protocol was approved by the University of Colorado Institutional Review Board.
Procedures were explained to each subject and written consent was obtained prior to participation.
Subjects' height and weight were recorded with a tape measure and scale, respectively.
Subjects' foot length, foot width, and arch length were measured with a Brannock device (The Brannock Device Company, Liverpool, NY) \citep{ASTM2017}.
Both foot length and arch length were measured in centimeters.
Foot width was measured as an ordinal size (e.g.~A, B, C, D, E), and then converted to a linear measurement in centimeters (The Brannock Device Company, Liverpool, NY).

\begin{figure}
\hypertarget{fig:testSetup}{%
\centering
\includegraphics[width=1\textwidth,height=\textheight]{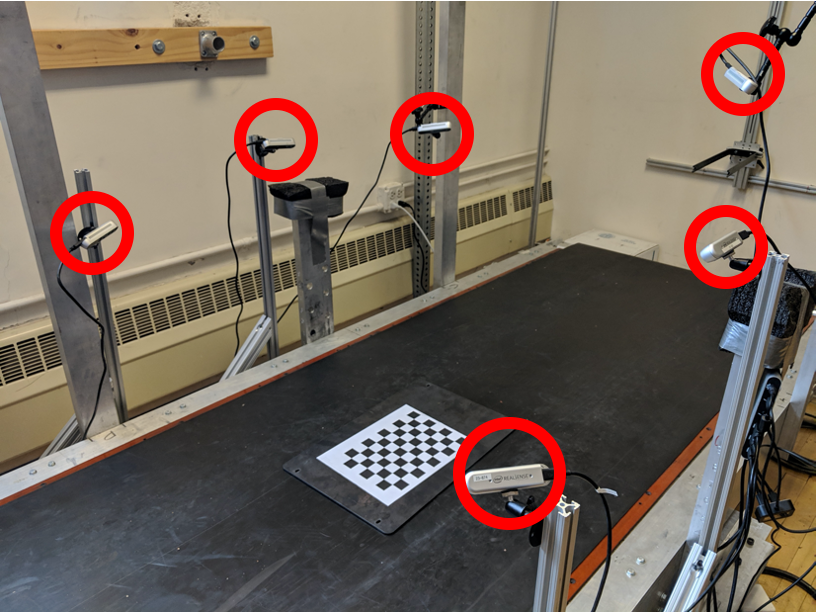}
\caption{Capture setup of 6 Intel RealSense D415 Depth Cameras (circled in red) placed around a treadmill. The checkerboard shown was used to calibrate the cameras using the DynaMo package.}\label{fig:testSetup}
}
\end{figure}

Six Intel RealSense D415 Depth Cameras (Intel, Santa Clara, CA) were placed and calibrated around a custom-built level treadmill in the University of Colorado Boulder Locomotion Laboratory, as shown in Fig.~\ref{fig:testSetup}.
The DynaMo software package was used to capture depth images of the right foot at 90 frames-per-second while subjects walked on the treadmill, and convert each frame's depth images to a single point cloud \citep{Boppana2019}.

The treadmill was set to an average walking pace of 1.4 m/s \citep{Browning2006}.
Reflective markers were placed on the subject's right foot and a black sock over their left foot to aid in right foot identification.
Subjects first walked for one minute to warm-up and fall into a natural cadence.
The operator then collected 10 seconds of data to capture approximately 10 steps.
The data were reviewed to ensure the subject stayed in frame from heel-strike to toe-off during capture. If needed, the subject's placement was shifted and data was collected again, up to two times.

\hypertarget{data-processing}{%
\subsection{Data Processing}\label{data-processing}}

\begin{figure}
\hypertarget{fig:dataflow}{%
\centering
\includegraphics[width=1\textwidth,height=\textheight]{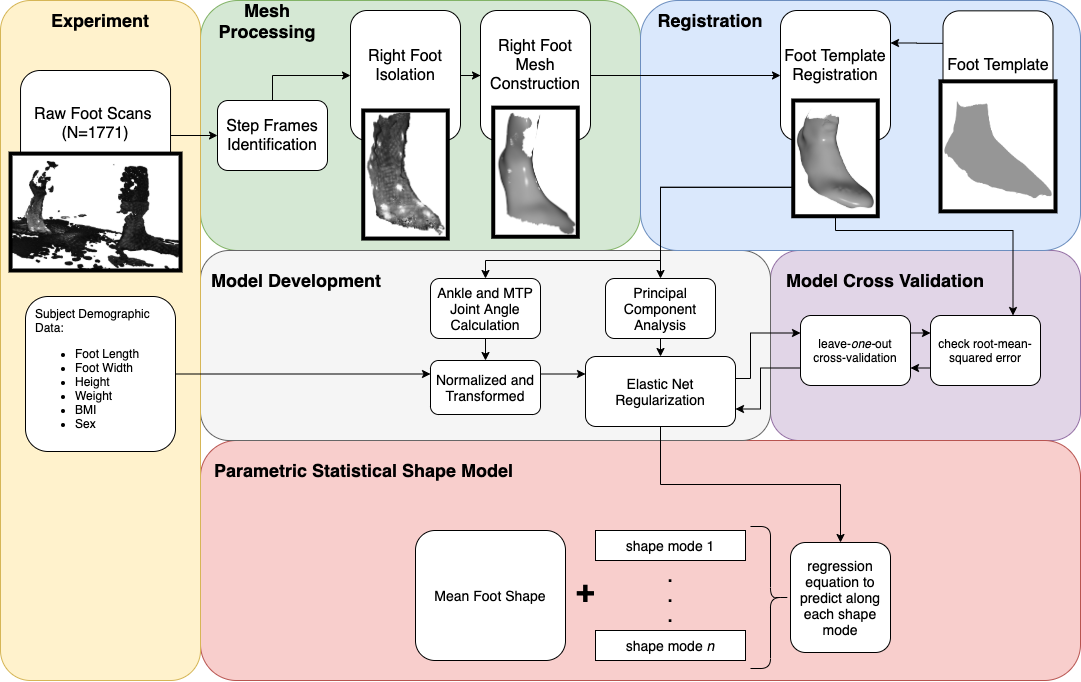}
\caption{Flowchart of processing steps for statistical shape model creation}\label{fig:dataflow}
}
\end{figure}

(Fig. \ref{fig:dataflow}) provides an overview of the data processing workflow.
The following paragraphs summarize the workflow, while more detail is provided in supplementary methods.

For each subject, a candidate heel-strike to toe-off event was manually identified across all captures by taking into account point cloud quality due to the high computational power required to process all heel-strike to toe-off events.
The depth images captured by each depth camera were processed into point clouds using the DynaMo package \citep{Boppana2019}.
From each point cloud, the right foot was isolated and transformed into a triangle mesh \citep{Rusu2011, Fischler1981, Bernardini1999, Zhou2018}.
Since every depth image was captured independently by the cameras, the amount and location of points which represented the foot were not consistent.
In addition, the captured data may have holes in the surface representing the foot.
Registration of all scans to a common template represents every scan by an equal number of points, and ensures any missing points are properly interpolated.
The right foot meshes were then iteratively registered using a three-step fitting process to an averaged high-quality static template scan from a previous study \citep{Reed2013}.
First scans were roughly aligned using a point-to-plane iterative-closest-point algorithm \citep{Chen1992}, implemented in Open3D \citep{Zhou2018}.
Next, the radial-basis function fitting algorithm from the GIAS2 software package \citep{Zhang2016} was run twice using a thin-plate spline to approximate the foot surface \citep{Park2015a, KIM2016}.
The mid-stance scan from each subject was registered first to the template, and then the registration process was run both forwards towards toe-off and backwards towards heel-strike, on a scan-by-scan basis, using the previously registered scan as a template for the next scan.
Accuracy was checked by comparing registered scans with the processed scans by finding corresponding points between both, and calculating the root-mean-squared error (RMSE) between the corresponding points.

Anatomical landmarks can be reliably approximated from the registered scans \citep{VandenHerrewegen2014b}.
The first metatarsal head, fifth metatarsal head, and second toe landmarks were used to align all scans to be centered at the second metatarsal head, with the forward axis pointing towards the second toe.
Landmarks around the metatarsal-phalangeal (MTP) joint and ankle joint were used to calculate ankle, MTP, and foot kinematics for each subject's scans with respect to the joint angles at the subject's mid-stance scan.
Relevant joint angles include dorsi/plantarflexion, ankle inversion/eversion, ankle internal/external rotation, MTP dorsi/plantarflexion, foot inversion/eversion, and foot internal/external rotation angles

\hypertarget{model-construction}{%
\subsection{Model Construction}\label{model-construction}}

Principal component (PC) analysis is a dimensionality-reduction method commonly in constructing SSMs \citep{Reed2008, Park2015a, Conrad2019, Stankovic2020}.
The first PC represents an axis containing the largest variance in the dataset, and each subsequent PC describes the largest variance orthogonal to the previous component's axis.
Therefore, PCs allow for a new, smaller set of orthogonal variables to be defined which represent the variance in the dataset.

Let \(N\) equal the number of total scans in the dataset, and \(n=29873\) equal the number of vertices in each registered scan. The scikit-learn module \citep{JMLR:v12:pedregosa11a} was used to incrementally calculate the maximum \(N\) PCs which represent the dataset.
Each scan in the dataset is represented in the PC model with \(N\) PC scores.
All PC scores are centered around 0, which represents the mean foot scan of the dataset containing all subjects.
Each PC represents a shape mode in the SSM, where each score represents a deviation from the mean foot along the shape mode axis. The resultant PC model can be used to inverse transform a vector of length \(N\) PC scores into a \(29873\times 3\) vector, which represents the location of the vertices in the foot shape. Not all PCs were retained in the model since the first few PCs explain a majority of the variance, while additional PCs may be accounting for noise.

Subject demographic data and calculated joint angles were incorporated into the SSM by developing multivariate linear regression models based on these features.
This was used to predict each PC score, which can then be inverse-transformed into a foot shape.
Subject demographic data and joint angles were normalized and power-transformed to aid in regression development \citep{Yeo2000}.
An elastic net regularization algorithm \citep{Zou2005} was run for each multivariate regression to calculate normalized feature coefficients for each PC score's regression.
Two different sets of predictors were created, one with all subject demographic data and calculated joint angles, and one with the highly cross-correlated predictors of arch length, body-mass index, and height were removed (see Supplementary Figures).
Six potential models were built as combinations between the number of PCs predicted which explained 95\%, 98\%, and 99.7\% of the variance, and the two predictor sets.

\hypertarget{model-validation}{%
\subsection{Model Validation}\label{model-validation}}

All six models were validated for performance using leave-one-out cross-validation, where scans from each subject were set as the validation set, and models were trained on the remaining dataset.
Model performance during validation was quantified with the root mean squared error (RMSE) of the predicted foot shape to the corresponding registered scan.
A two-way RMANOVA analysis was run on the error distributions to test the effect of constructing a predictor with the different number of PCs, and between using the two variable sets.
The chosen model was retrained on the whole dataset before being analyzed.
\# Results

\begin{figure}
\hypertarget{fig:scans}{%
\centering
\includegraphics[width=1\textwidth,height=\textheight]{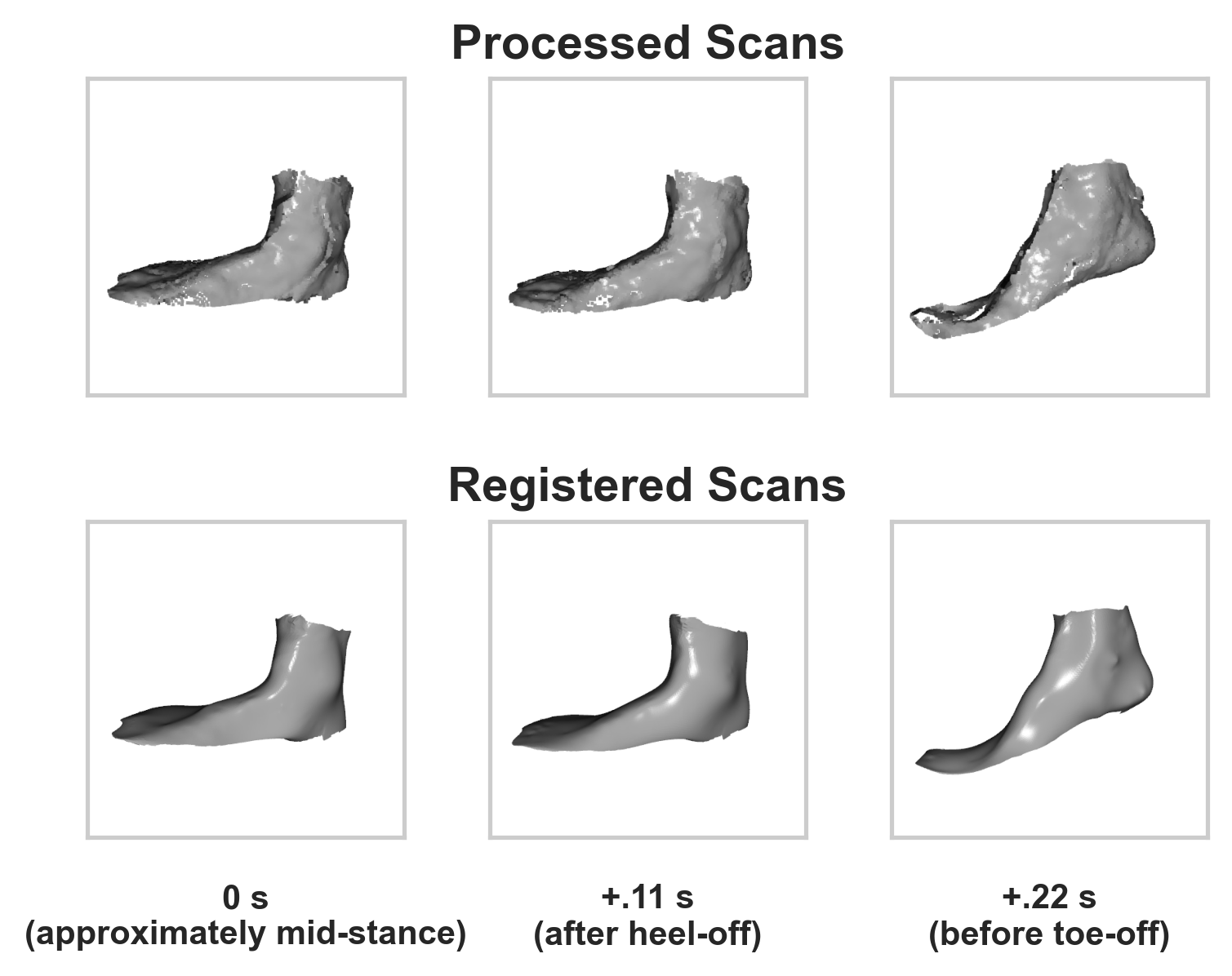}
\caption{Processed and registered scans of one subject during heel-off, shown 10 frames (.11 seconds) apart}\label{fig:scans}
}
\end{figure}

A total of 1771 scans were analyzed across all 30 subjects.
Each subject's stance phase ranged from 52-69 scans (mean=59).
(Fig. \ref{fig:scans}) shows a set of raw and registered scans from one subject.
All processed scans were registered to the template with a median registration accuracy of 1.0 \(\pm\) 0.6 mm.

\begin{figure}
\hypertarget{fig:modelperf}{%
\centering
\includegraphics[width=1\textwidth,height=\textheight]{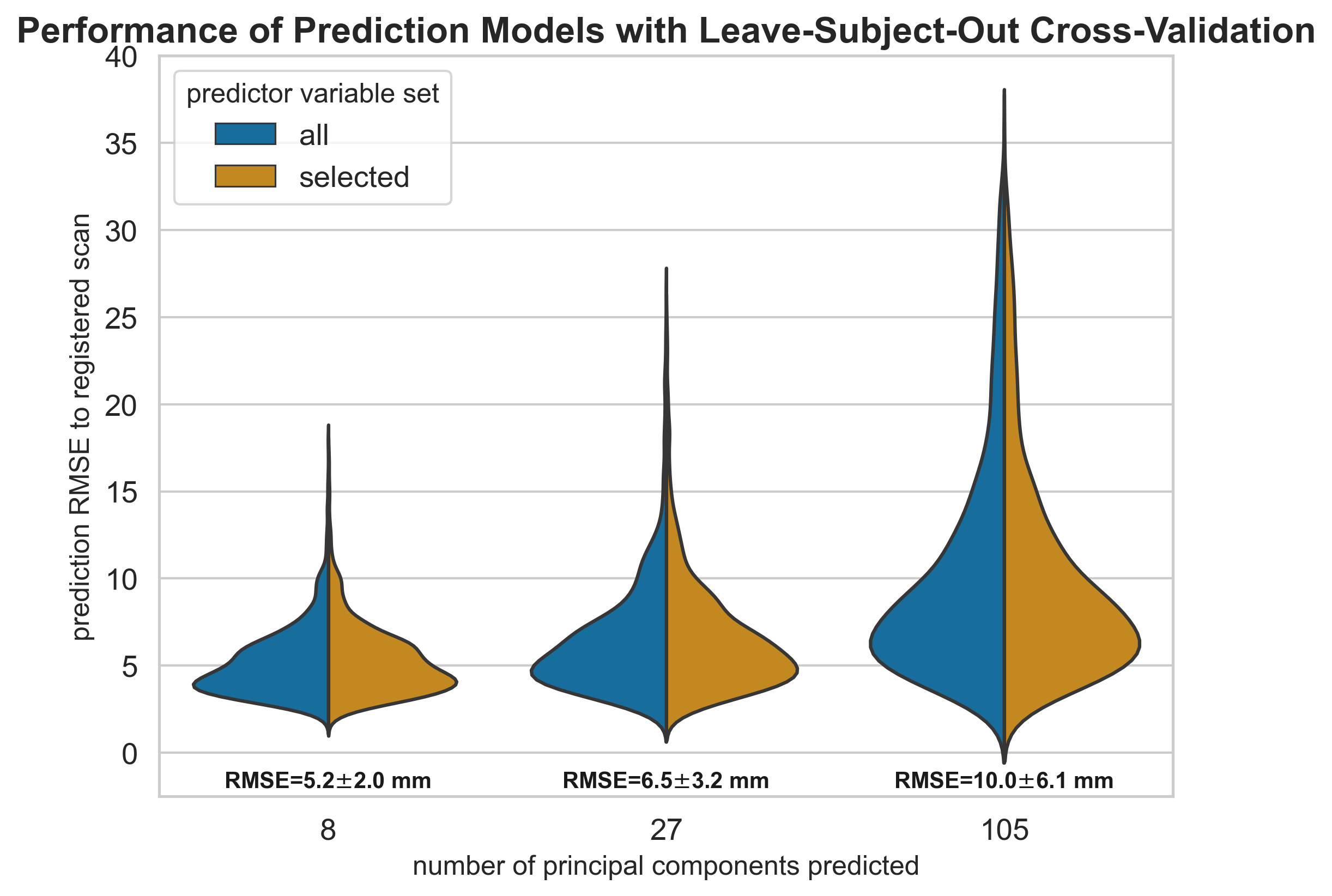}
\caption{Distribution of errors across the various prediction models leave-subject-out cross-validation results. Model RMSE mean and standard deviation are shown above each distribution}\label{fig:modelperf}
}
\end{figure}

The PCA analysis of all registered scans found the first 8 PCs to represent approximately 95\% of the variance, the first 27 PCs to represent approximately 98\% of the variance, and the first 105 PCs to represent approximately 99.7\% of the variance.
(Fig. \ref{fig:modelperf}) shows the distribution of cross-validation RMSEs for each of the six elastic net regression models tested.
RMSE distributions did not meet assumptions for normality, but RMANOVA was still used to compare models due to its resiliency to deviations from normality.
A significant difference was found between predicting different numbers of PCs (F=1595.0, p\textless0.001), predicting between the two variable sets (F=81.6, p\textless0.001), and the interaction between both factors (F=213.7, p\textless0.001).
Significant differences were found between all three levels of the predicted number of PCs (p-adj\textless0.001) with a Tukey post-hoc HSD test.
No significant difference was found between the two variable sets (p-adj=0.42).
Therefore, the model predicting 8 PCs with the selected variable set was chosen for its simplicity and performance.

\begin{figure}
\hypertarget{fig:coefs}{%
\centering
\includegraphics[width=1\textwidth,height=\textheight]{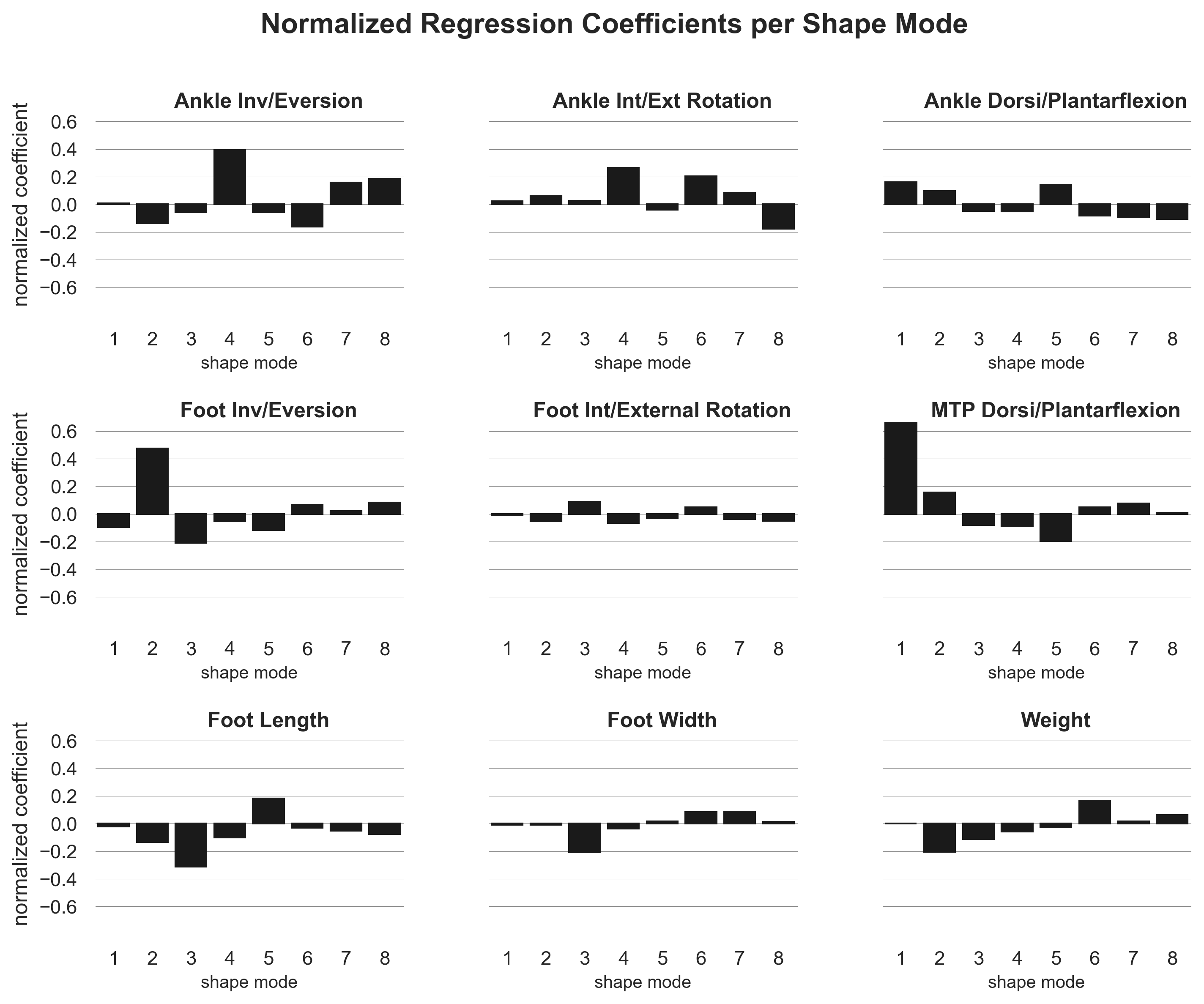}
\caption{Each graph represents the predictor's effects on the shape mode by visualizing the model's normalized coefficients. Larger absolute values indicate a larger effect from the predictor on the shape mode.}\label{fig:coefs}
}
\end{figure}

Each retained PC is a shape mode in the model. (Fig. \ref{fig:coefs}) shows the chosen model's normalized regression coefficient values for each shape mode.
The coefficients for the sex predictor are not shown as they were calculated to be zero for every shape mode.

\begin{figure}
\hypertarget{fig:pca_quad}{%
\centering
\includegraphics[width=1\textwidth,height=\textheight]{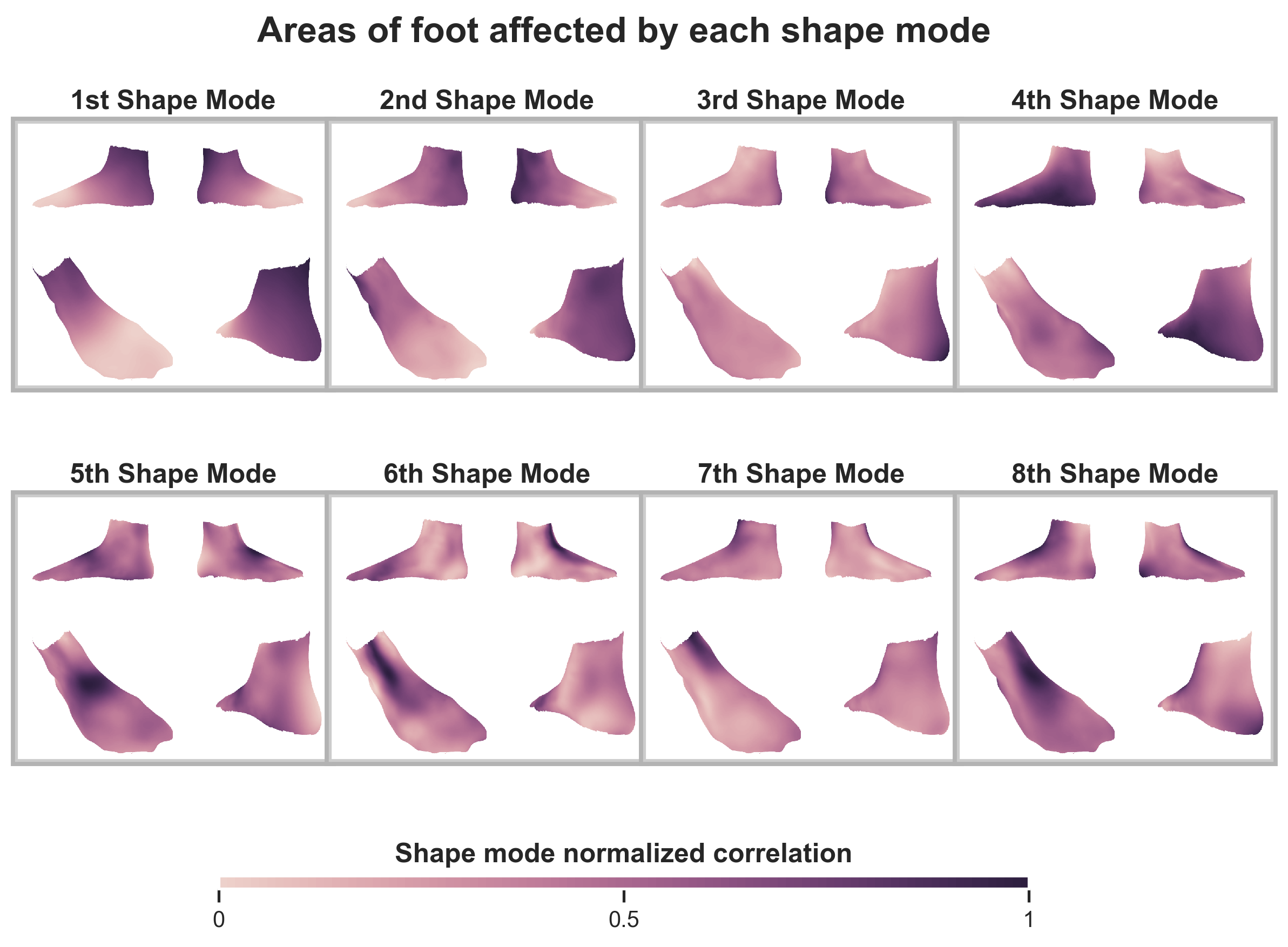}
\caption{Each shape mode's principal axis represented as a heatmap overlaid on the mean foot and shown from 4 different point-of-views. The darker regions represent vertices which are most correlated with the shape mode's principal axis, and therefore see deformations in the shape mode.}\label{fig:pca_quad}
}
\end{figure}

\begin{figure}
\hypertarget{fig:pca_overlay}{%
\centering
\includegraphics[width=1\textwidth,height=\textheight]{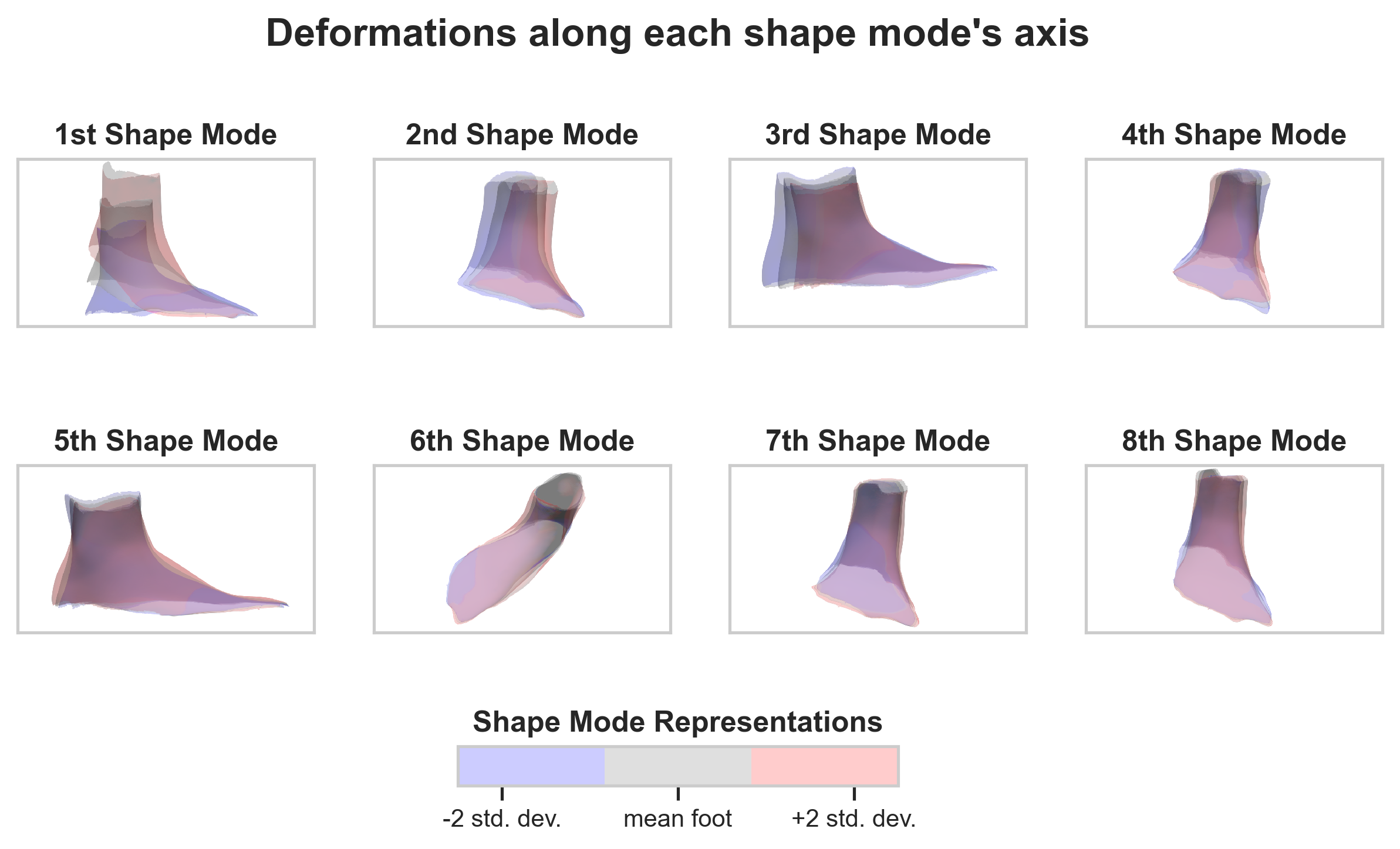}
\caption{Foot shape deformation at +2 and -2 standard deviations along each shape mode's principal axis, overlaid on the mean foot. The point-of-view is set to highlight the major variance along each shape mode's axis.}\label{fig:pca_overlay}
}
\end{figure}

(Fig. \ref{fig:pca_quad}) shows each shape mode's axis represented on the mean foot, highlighting which areas of the foot are affected by deformations in each shape mode.
(Fig. \ref{fig:pca_overlay}) shows the \(\pm\) 2 standard deviations of deformation along each shape mode overlaid on the mean foot.
Supplementary information includes correlation between figures, ratio of total variance each retained PC accounts for, and a video showing the predictive capability of the model.

\hypertarget{discussion}{%
\section{Discussion}\label{discussion}}

This study was designed to construct and evaluate a parametric SSM in explaining and predicting dynamic foot morphology changes across the subject population.
The model was able to predict dynamic foot shape across the subject population with an average RMSE of 5.2 \(\pm\) 2.0 mm. For context, if all possible prediction error was accumulated to only affect length and width, it would be higher than the half-size step of the American shoe sizing system \citep{Luximon2013}, but less than inter-brand variability of shoe length and shoe width \citep{Wannop2019}.
Further, this error is lower than the RMSEs of other parametric SSMs that predicted static standing child body shape (mean=10.4mm) \citep{Park2015a}, dynamic shoulder deformation (mean=11.98mm) \citep{KIM2016} and child torso shape (mean=9.5mm) \citep{Park2017}. Note though, that the presented model may have lower prediction errors due to the foot being a relatively smaller section of the body to model. Grant et al's model reconstructed internal foot bones with much lower RMSEs from sparse anatomical landmarks (1.21-1.66 mm for various foot segments) \citep{Grant2020} but was trained with higher resolution MRI images. Other efforts to create statistical foot shape models did not incorporate parametric prediction of foot shape \citep{Conrad2019, Stankovic2020}.

The first, second, and fourth shape modes, accounting for a total of 86.7\% of total variance, capture gross foot motion.
Foot motion during stance is dominated by MTP and ankle dorsi/plantarflexion \citep{Leardini2007}, which is captured in the first shape mode (Fig.~\ref{fig:pca_overlay}).
The second and fourth shape modes capture gross changes in foot rotation from frontal and transverse plane movements at the MTP and ankle joints, respectively (Fig.~\ref{fig:pca_overlay}).
The second shape mode is most affected by foot inversion/everison around the MTP joint.
The second shape mode also captures girth scaling at the ankle joint, as seen in (Fig.~\ref{fig:pca_overlay}) by how the ankle girth decreases along the axis, and is affected by weight (Fig.~\ref{fig:coefs}).
The fourth shape mode is affected by ankle inversion/eversion and internal/external rotation.
Foot inversion/eversion, ankle inversion/eversion, and ankle internal/external rotation are expected to vary across the stance phase (\citep{Leardini2007}), which leads to the observed changes in gross movement.
However, the second and fourth shape modes are slightly affected by foot length, which may suggest inter-individual effects in foot inversion/eversion, ankle inversion/eversion, and internal/external rotation during gait.
There is a slight correlation between these angles and foot length (see supplementary figures), which may be due to differences in cadence when walking at the treadmill's set speed.
Individuals were given time to acclimate to the treadmill's set speed, but the speed may not have been their preferred walking speed.

The third shape mode captures foot shape scaling at the rearfoot, as highlighted in (Fig.~\ref{fig:pca_quad}).
Foot length shrinks when moving positively along the third shape mode (Fig.~\ref{fig:pca_overlay}), and thus has a negative effect from foot length.
There are also negative effects from foot width and weight, which may be due to their correlation to foot length (see supplementary figures).
Rearfoot morphology along this shape mode has a more rounded shape in the negative direction, and a sharper shape in the positive direction (Fig.~\ref{fig:pca_overlay}).
There is also a negative effect from foot inversion/eversion (Fig.~\ref{fig:coefs}), indicating that with foot eversion, a sharper rearfoot shape is expected.
This may be due to foot eversion at heel-off \citep{Leardini2007}, where the foot unloads from a rounder weight-bearing rearfoot to a sharper non-weight bearing rearfoot shape.

Midfoot girth increases and the rearfoot is rounder along the fifth shape mode's axis (Fig.~\ref{fig:pca_overlay}).
The fifth shape mode is positively affected by foot length and negatively by MTP dorsi/plantarflexion (Fig.~\ref{fig:coefs}).
This suggests that static midfoot girth increases with foot length, and decreases through heel-off as the MTP dorsiflexes.
Rearfoot morphology is rounder for longer foot lengths but gets sharper through heel-off with MTP dorsiflexion, much like in the third shape mode.
Midfoot girth was previously found to decrease during stance phase compared to statically standing \citep{Grau2018}, most likely due to intrinsic and extrinsic foot muscle contraction \citep{Scott1993, Gefen2000}.
However, it was not noted where during stance phase midfoot girth decreases, but it can now be assumed it occurs during heel-off.

The sixth shape mode captures girth changes at the ankle, midfoot, and the medial MTP joint region (Fig.~\ref{fig:pca_quad}), with girth increasing along the axis.
There are positive effects from ankle internal/external rotation and weight, while there is a negative effect from ankle inversion/eversion (Fig. \ref{fig:coefs}).
Static MTP, midfoot, and ankle girth may therefore increase with subject weight.
Dynamic girth changes in these regions may occur as the ankle everts and internally rotates just prior to toe-off, where muscle activation is needed to push the foot off the ground.
The foot is stiffened through tension in the MTP joints in order to prepare for toe-off \citep{Hicks1954}, and the MTP joints are known to move relatively within the foot during gait \citep{Wolf2008, Lundgren2008} which may be resulting in the increased girth at the MTP joint.
A similar mechanism may be occuring at the ankle joint during ankle inversion and internal rotation, where tension from muscle activation prior to toe-off may cause increased girth.

The seventh and eight shape modes, accounting for 1.3\% of total variance, capture girth increases near the medial malleolus along their axes (Fig.~\ref{fig:pca_quad}).
They are both positively affected by ankle inversion/eversion (Fig.~\ref{fig:coefs}), and the eight shape mode is further negatively affected by ankle internal/external rotation.
This may suggest that the girth around the medial malleolus decreases prior to push-off, as the ankle everts and internally rotates.

Observed girth changes at the ankle joint, medial malleolus, midfoot, and MTP joint can be directly mapped to footwear design recommendations for increased fit and comfort. Midfoot girth decreased as the MTP joint is dorsiflexing after heel-off.
Midfoot, ankle, and MTP joint girth increased and medial malleolus girth decreased through ankle eversion and external rotation just prior to toe-off.
Footwear should be designed to follow these volume changes as the footwear itself goes through the same motions, to ensure proper support for the foot to drive the footwear through the stance phase and toe-off.
For example, footwear may be designed to first contract as the MTP joint dorsiflexes, then subsequently expand around the midfoot, ankle and MTP joints while contracting around the medial malleolus as the ankle everts and externally rotates.

A number of limitations in this study should be noted.
The elastic-net method is able to retain cross-correlated predictors, but still requires some bias in the dataset to predict scenarios where cross-correlated predictors are independent \citep{Zou2005}.
Therefore, the presented model may not be valid for predicting changes in morphology due to independent changes in joint angles outside of stance phase, or for variance in foot width or weight compared to foot length not captured in the subject population.

The model did not capture differences between male and female feet.
Studies found that sex differences in foot shape after scaling for foot length were not significant \citep{Kouchi2009, Barisch-Fritz2014a, Conrad2019}, or were small in magnitude \citep{Wunderlich2001, Krauss2008}.
No subject demographic data was collected to account for differences in foot shape due to ethnicity \citep{Jurca2019}.
No data was captured on the foot's plantar surface due to limitations with the scanning system; therefore foot arch changes were not captured.
Data captured around the toes had high noise, which necessitated smoothing the toes in the template to ease fitting.
Future advances in 4D scanning may alleviate some of these concerns, and also allow for expansion of this model to higher frequency foot motions, such as running.

\hypertarget{conclusions}{%
\section{Conclusions}\label{conclusions}}

To the authors' knowledge, this is the first parametric foot SSM that captures and reconstructs dynamic motion.
The model was able to identity specific changes in foot morphology as they related to subject and kinematic parameters, and suggest footwear design techniques to increase fit and comfort.
The model is able to reconstruct a full 3D model when parameter values are provided, which offers shoe and last designers a design starting point, and the ability to test their designs on a range of subjects throughout stance phase.
\newpage

\hypertarget{figures-and-tables}{%
\section*{Figures and Tables}\label{figures-and-tables}}
\addcontentsline{toc}{section}{Figures and Tables}

All tables, figures, and respective captions are listed below

\hypertarget{tbl:groups}{}
\begin{longtable}[]{@{}llll@{}}
\caption{\label{tbl:groups}Enrollment groups based on reported height. 5 subjects were enrolled in each group}\tabularnewline
\toprule
\begin{minipage}[b]{0.06\columnwidth}\raggedright
Sex\strut
\end{minipage} & \begin{minipage}[b]{0.27\columnwidth}\raggedright
5th-35th percentile Height\strut
\end{minipage} & \begin{minipage}[b]{0.28\columnwidth}\raggedright
35th-65th percentile Height\strut
\end{minipage} & \begin{minipage}[b]{0.28\columnwidth}\raggedright
65th-95th percentile Height\strut
\end{minipage}\tabularnewline
\midrule
\endfirsthead
\toprule
\begin{minipage}[b]{0.06\columnwidth}\raggedright
Sex\strut
\end{minipage} & \begin{minipage}[b]{0.27\columnwidth}\raggedright
5th-35th percentile Height\strut
\end{minipage} & \begin{minipage}[b]{0.28\columnwidth}\raggedright
35th-65th percentile Height\strut
\end{minipage} & \begin{minipage}[b]{0.28\columnwidth}\raggedright
65th-95th percentile Height\strut
\end{minipage}\tabularnewline
\midrule
\endhead
\begin{minipage}[t]{0.06\columnwidth}\raggedright
Female\strut
\end{minipage} & \begin{minipage}[t]{0.27\columnwidth}\raggedright
4'11``-5'3''\strut
\end{minipage} & \begin{minipage}[t]{0.28\columnwidth}\raggedright
5'3``-5'5''\strut
\end{minipage} & \begin{minipage}[t]{0.28\columnwidth}\raggedright
5'5``-5'8''\strut
\end{minipage}\tabularnewline
\begin{minipage}[t]{0.06\columnwidth}\raggedright
Male\strut
\end{minipage} & \begin{minipage}[t]{0.27\columnwidth}\raggedright
5'4``-5'8''\strut
\end{minipage} & \begin{minipage}[t]{0.28\columnwidth}\raggedright
5'8``-5'11''\strut
\end{minipage} & \begin{minipage}[t]{0.28\columnwidth}\raggedright
5'11``-6'2''\strut
\end{minipage}\tabularnewline
\bottomrule
\end{longtable}

\newpage

\newpage

\newpage

\newpage

\newpage

\newpage

\newpage

\newpage
\subsubsection*{Acknowledgements}
{\small The authors would like to thank Rodger Kram for providing the laboratory space and treadmill used in the study, Wouter Hoogkamer for assistance with equipment setup, and Steven Priddy for assistance isolating the right foot from the 4D scans. The authors would also like to thank Brian Corner and Matthew Reed for providing a high-quality averaged foot-scan to be used as the template for registration. This project was supported with a National Science Foundation Graduate Research Fellowship Grant DGE 1650115.}

\bibliography{references}
\begin{center}
\includegraphics[page=1,width=\linewidth,height=\paperheight]{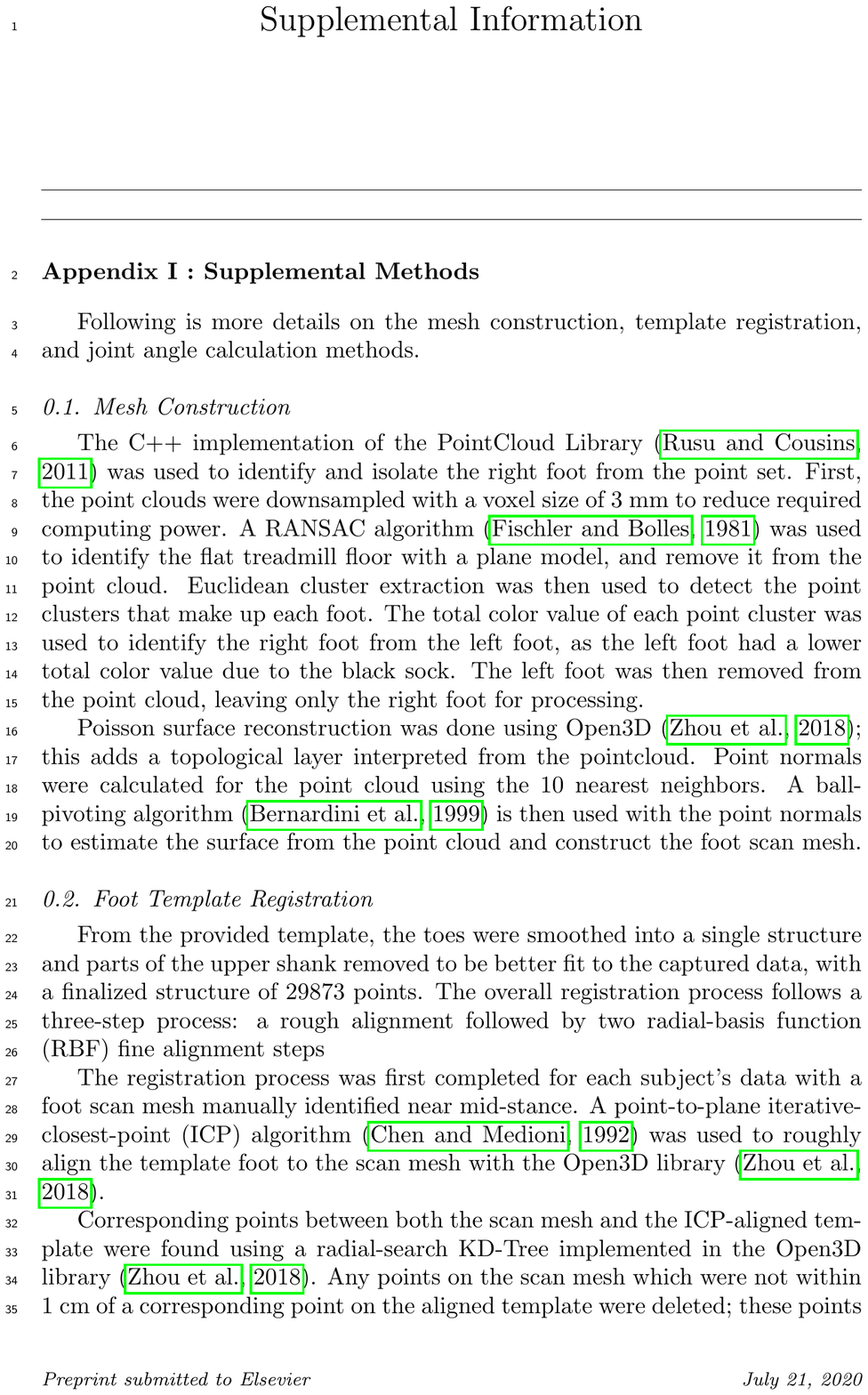} % 
\includegraphics[page=2,width=\linewidth,height=\paperheight]{FootMorphSupp.pdf} % 
\includegraphics[page=3,width=\linewidth,height=\paperheight]{FootMorphSupp.pdf} % 
\includegraphics[page=4,width=\linewidth,height=\paperheight]{FootMorphSupp.pdf} % 
\includegraphics[page=5,width=\linewidth,height=\paperheight]{FootMorphSupp.pdf} % 
\includegraphics[page=6,width=\linewidth,height=\paperheight]{FootMorphSupp.pdf} % 

\end{center}
\end{document}